\documentclass[twocolumn]{aastex701}

\usepackage[utf8]{inputenc}       
\usepackage[version=4]{mhchem}    

\begin{document}

\title{Perihelion Asymmetry in the Water Production Rate of the Interstellar Object 3I/ATLAS}

\author[orcid=0000-0002-6144-3062, gname=Hanjie, sname=Tan]{Hanjie Tan}
\affiliation{Planetary Environmental and Astrobiological Research Laboratory (PEARL), School of Atmospheric Sciences, Sun Yat-sen University, Zhuhai, China}
\email{h.tan1996@google.com}

\author[orcid=0000-0001-6241-5744, gname=Xiaoran, sname=Yan]{Xiaoran Yan}
\affiliation{National Research Council, Institute of Applied Physics "Nello Carrara" (IFAC–CNR), 50019 FI, Italy}
\email[show]{yanxr159@gmail.com}

\author[orcid=0000-0003-3841-9977, gname=Jian-Yang,sname=Li]{Jian-Yang Li}
\affiliation{Planetary Environmental and Astrobiological Research Laboratory (PEARL), School of Atmospheric Sciences, Sun Yat-sen University, Zhuhai, China}
\affiliation{Xinjiang Astronomical Observatory, Chinese Academy of Sciences, Urumqi, China}
\email[show]{lijianyang@mail.sysu.edu.cn}

\begin{abstract}

3I/ATLAS is an interstellar object whose activity provides critical insights into its composition and origin. However, due to its orbital geometry, the object is too close to the Sun near perihelion to be observed from the ground, and space-based measurements are therefore required. Here we characterize the water production rate of 3I/ATLAS using SOHO/SWAN Lyman-$\alpha$ observations from 2025 November to December (heliocentric distances 1.4 to 2.2 au) with 3D Monte Carlo modeling. We report a peak post-perihelion water production rate of $Q_{\mathrm{H_2O}} \approx 4 \times 10^{28}$ molecules~s$^{-1}$, corresponding to a minimum active fraction of $\sim$30\% (assuming a maximum nucleus radius of 2.8 km). Comparison of our post-perihelion measurements with published pre-perihelion results reveals a heliocentric asymmetry, with an $r^{-5.9 \pm 0.8}$ scaling for the inbound rise, followed by a shallower $r^{-3.3 \pm 0.3}$ scaling during the outbound decline, where $r$ is heliocentric distance. The post-perihelion behavior indicates that the water production of 3I/ATLAS was driven primarily by the varying solar insolation acting on a stable active area. Combined with other evidence, including comparison with the hyperactive comet 103P/Hartley 2, our findings suggest that its water production is likely dominated by a distributed source of icy grains. Furthermore, it displayed remarkable stability in the activity with no signs of outbursts or rapid depletion of water production. 


\end{abstract}




\section{Introduction} \label{sec:intro}

\begin{figure*}[ht!]
    \centering
    \includegraphics[width=0.9\linewidth]{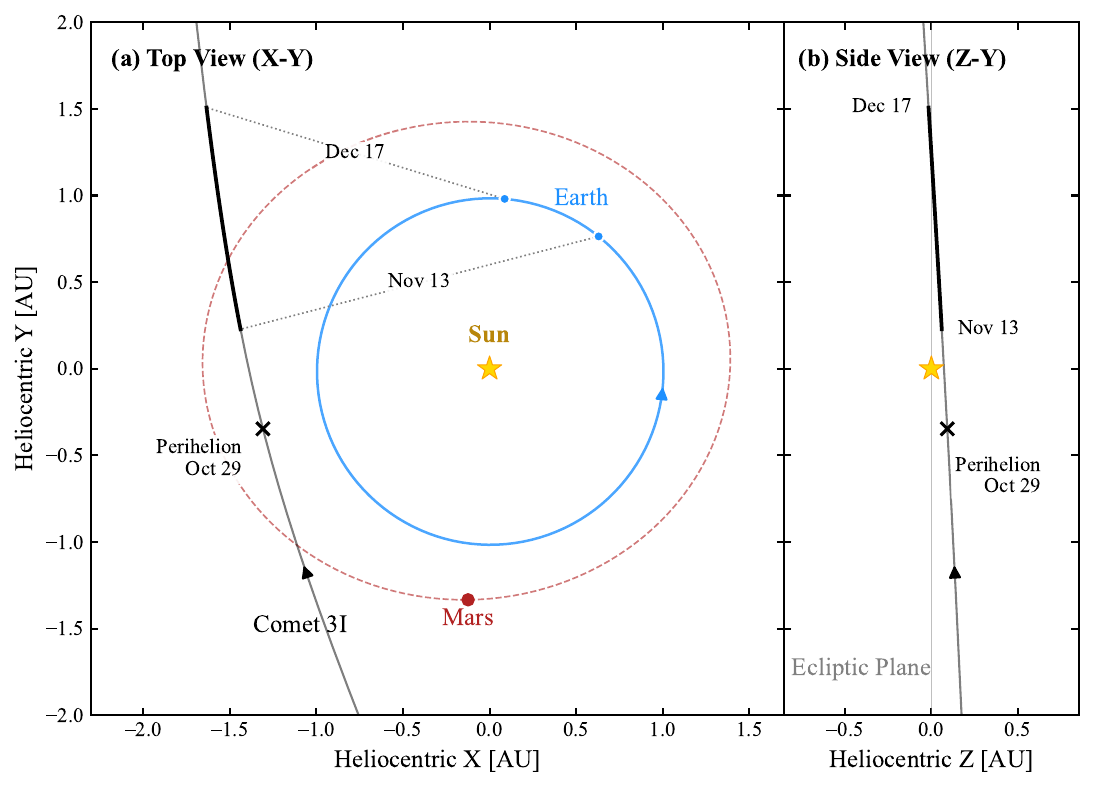}
    \caption{Orbital geometry of 3I/ATLAS during the SWAN observing window. The low-inclination trajectory is shown relative to the ecliptic plane, with Sun, Earth and Mars orbits provided for scale.}
    \label{fig:orbit}
\end{figure*}

The discovery of 3I/ATLAS expands the sparse inventory of known Interstellar Objects. In contrast to the inactive 1I/`Oumuamua \citep{Hui2019} or the volatile-rich 2I/Borisov \citep{Bodewits2020, Yang2021}, 3I/ATLAS provides a unique opportunity to monitor the thermal evolution of a pristine interstellar nucleus under increasing solar insolation. Its high hyperbolic excess velocity ($v_{\infty} \approx 58$ km s$^{-1}$) implies a dynamical age of 3 to 11 Gyr \citep{Taylor2025}, thereby providing constraints on planetesimal formation mechanisms in the early history of the Galaxy.


Early characterization of 3I/ATLAS at large heliocentric distances ($r > 3$~au) indicated a coma driven by super-volatiles, specifically CO and CO$_2$, with a high CO$_2$/H$_2$O abundance ratio \citep{Cordiner2025}. The object’s overall brightening followed a moderate heliocentric index ($n = 3.8 \pm 0.3$), consistent with activity driven by super-volatile sublimation \citep{Ye2025, Jewitt2025b}. As the comet crossed the water-ice sublimation line ($r \approx 2$ to $3$~au), its activity evolved significantly. The pre-perihelion light curve at $r < 2$~au exhibited a steep brightening trend, following a power-law dependence of $r^{-7.5 \pm 1.0}$ \citep{Zhang2025}. This steep slope implies a transition in the activity driver, likely associated with the onset of significant water-ice sublimation.

During perihelion passage, comets are exposed to maximum solar insolation, typically resulting in peak activity. For instance, the interstellar comet 2I/Borisov exhibited an optical outburst followed by a rapid decline in water production near perihelion \citep{Jewitt2020, Xing2020}, indicating a significant transition in its activity state. Consequently, the post-perihelion evolution of 3I/ATLAS offers a unique opportunity to constrain the impact of thermal processing on an interstellar nucleus. However, due to the specific orbital geometry of 3I/ATLAS, this critical phase coincided with small solar elongations, precluding ground-based observations.

In this Letter, we present post-perihelion water production rates derived from space-based SOHO/SWAN Lyman-$\alpha$ observations. By combining these results with the pre-perihelion measurements, we investigate the heliocentric evolution of the water production rate. We report a pronounced perihelion asymmetry and a sustained level of high activity post-perihelion, consistent with a highly active comet \citep{Combi2011b} primarily driven by sublimation from a distributed source of icy grains.

\section{Observations and Data Reduction}
\label{sec:observations}

\begin{figure*}[ht!]
    \centering
    \includegraphics[width=\linewidth]{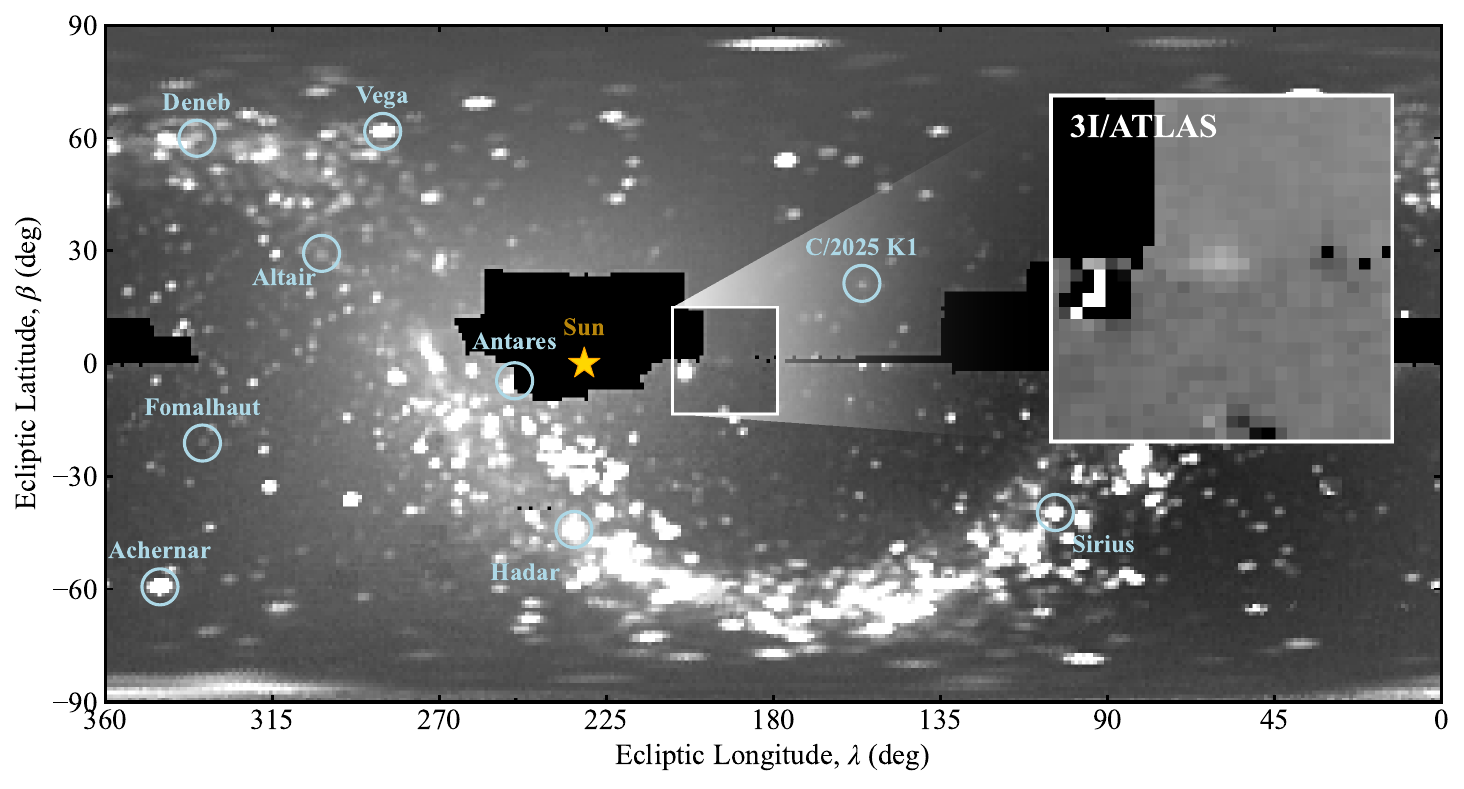}    
    \caption{Representative SWAN full-sky Lyman-$\alpha$ map (2025 Nov 13), showing the simultaneous detection of comet C/2025 K1 (ATLAS) and numerous UV-bright stars. The zoomed inset displays the differential image centered on 3I/ATLAS, illustrating its stellar-like appearance after the pipeline subtracts the IPH background and faint stellar sources. Residual artifacts remain for luminous stars.}    
    \label{fig:swan_map}
\end{figure*}

\subsection{SOHO/SWAN Observations}

We analyzed the hydrogen coma of 3I/ATLAS using the Solar Wind ANisotropies (SWAN) instrument on board the Solar and Heliospheric Observatory (SOHO) \citep{Bertaux1995, Domingo1995}. The instrument consists of two sensor units that scan the northern and southern ecliptic hemispheres to produce daily full-sky maps with a spatial resolution of $1^{\circ}$. These images measure solar Lyman-$\alpha$ photons resonantly scattered by neutral hydrogen atoms \citep{Bertaux1998, Makinen2001a}. The effective epoch for each daily map is defined as the midpoint of the $\sim$24-hour integration sequence. As shown in Figure \ref{fig:orbit}, our observations cover the post-perihelion period from 2025 November 13 to December 17, corresponding to a heliocentric distance range of $r \approx 1.4$ to 2.2~au.

To ensure the accuracy of our photometric analysis, we applied a stringent quality-filtering procedure to minimize contamination from the variable background (Figure~\ref{fig:swan_map}). The Lyman-$\alpha$ background is dominated by emission from interplanetary hydrogen (IPH) and numerous ultraviolet (UV) bright stars. Consequently, we excluded observations compromised by instrumental noise, contamination from bright stellar sources, or obstruction by the SOHO spacecraft structure. This selection process resulted in 16 images suitable for the analysis of water production rates.

Cross-calibration between SWAN and the PHEBUS instrument on BepiColombo indicates a true initial sensitivity closer to 3.2~R (counts s$^{-1}$)$^{-1}$ \citep{Pryor2024}, in contrast to the previously adopted standard of 4.1~R (counts s$^{-1}$)$^{-1}$ \citep{Quemerais2002}. Accordingly, we applied a factor of 1.28 to recalibrate the selected Level-2 full-sky maps, while adopting a conservative systematic uncertainty of 25\% for this absolute calibration.

To extract the cometary Lyman-$\alpha$ emission from the IPH background and stellar contamination, we employed a temporal differencing pipeline. Given the low ecliptic latitude of 3I/ATLAS, the comet was simultaneously observed by both SWAN sensor units. Consequently, the data from both sensors were reduced independently to account for their individual instrumental sensitivities \citep{Combi2011a}.

\subsection{Monte Carlo Particle Trajectory Model (MCPTM)}
\label{subsec:mcptm}

Water production rates were derived from the observed Lyman-$\alpha$ intensity maps using a three-dimensional, time-dependent Monte Carlo Particle Trajectory Model (MCPTM) \citep{Combi1988a, Combi1988b}. This model implements the vectorial formalism of \citet{Festou1981} to account for the time-varying observing geometry and the dissociation of the H atoms. The simulation integrates the trajectories of individual atoms under the combined influence of solar gravity and radiation pressure.

The model simulates the standard two-step photodissociation chain: $\mathrm{H_2O} \rightarrow \mathrm{OH+H} \rightarrow \mathrm{O+2H}$. Parent $\mathrm{H_2O}$ molecules are assumed to be released isotropically from the nucleus with an initial outflow velocity of $v_{\mathrm{H_2O}} = 0.85$~km~s$^{-1}$ \citep{Combi2004}. Upon dissociation, the daughter products are ejected isotropically in the rest frame of the parent. The H atoms produced from the first dissociation step ($\mathrm{H_2O} \rightarrow \mathrm{OH}$) receive an excess velocity of $v_{\mathrm{1}} = 20.0$~km~s$^{-1}$. For the second step ($\mathrm{OH} \rightarrow \mathrm{O}$), the H atoms receive an excess velocity of $v_{\mathrm{2}} = 8.0$~km~s$^{-1}$, while the OH radical has a velocity of $1.2$~km~s$^{-1}$. We adopted photochemical lifetimes at 1~au of $\tau_{\mathrm{H_2O}} = 8.2 \times 10^4$~s, $\tau_{\mathrm{OH}} = 2.25 \times 10^5$~s, and $\tau_{\mathrm{H}} = 1.5 \times 10^6$~s, scaled by $r^2$ \citep{Combi1988b, Bertaux1998, Makinen2001a}. A 10\% uncertainty was assumed for both the photodissociation rates and the velocity distributions in our error analysis.

The fluorescence efficiency ($g$-factor) was calculated as a function of the comet's heliocentric radial velocity to account for the Swings effect, utilizing the high-resolution solar Lyman-$\alpha$ line profile \citep{Lemaire1998, Lemaire2015}. The derived $g$-factors were scaled to the daily integrated solar Lyman-$\alpha$ irradiance obtained from the LISIRD database \citep{Kretzschmar2018}. A correction was applied for the difference in heliographic longitude between the Earth and the comet \citep{Makinen2001a}.

The water production rate $Q_{\mathrm{H_2O}}$ was determined by scaling the observed Lyman-$\alpha$ integrated flux to the MCPTM predictions. Using the model described by \citet{Combi1988b}, the relationship is expressed as:

\begin{equation}
Q_{\mathrm{H_2O}} = Q_{\mathrm{sim}} \, \frac{4\pi \, \Delta^2 \, \mathcal{F}_{\mathrm{obs}}}{\kappa \, g(r) \, \mathcal{N}_{\mathrm{sim}}(v, \tau)}
\end{equation}

where $\mathcal{F}_{\mathrm{obs}}$ is the integrated flux within the photometric aperture, $\Delta$ is the SOHO-comet distance, and $\kappa = 1.28$ is the calibration factor (Section~\ref{sec:observations}). $Q_{\mathrm{sim}}$ is the input production rate used in the Monte Carlo simulation, and $\mathcal{N}_{\mathrm{sim}}(v, \tau)$ represents the modeled number of hydrogen atoms within the aperture, calculated based on the adopted velocity field ($v$) and lifetimes ($\tau$).

We adopted an iterative predictor-corrector approach to solve for the water production rates, accounting for the heliocentric dependence of both the water production history and particle lifetimes. The calculation was repeated until the derived production rates converged.

The total uncertainty in the water production rate was estimated by propagating the individual error contributions in quadrature. The error includes: (1) a 25\% systematic uncertainty from the SOHO/SWAN absolute calibration; (2) a 10\% uncertainty in the $g$-factor arising from the solar flux calibration \citep{Woods2000, Shinnaka2017}; (3) a 10\% uncertainty assigned to both photochemical lifetimes and velocity distributions; (4) model systematics related to aperture effects; and (5) statistical noise in the measurements. The resulting total uncertainty, typically ranging from 30\% to 40\%, is consistent with the error budgets reported in previous SWAN analyses \citep{Bertaux2014, Combi2019}.

To validate our data reduction and modeling procedures, the absolute flux calibration, and model parameterizations, we applied our procedure to a set of well-characterized comets and compared our results with the literature. As detailed in Appendix~\ref{sec:appendix_validation}, our derived water production rates show good agreement with published datasets for 41P/Tuttle-Giacobini-Kresák \citep{Bodewits2018, Combi2019, Lis2019, Moulane2018}, 46P/Wirtanen \citep{Combi2020, Knight2021, Moulane2023}, and C/2020 F3 (NEOWISE) \citep{Aravind2025, Combi2021, Drozdovskaya2023, Faggi2021}. This consistency holds for both the absolute production levels and their heliocentric dependence. The successful recovery of production rates across comets of diverse dynamical classes and activity levels confirms the accuracy of our analysis and supports the reliability of the results presented for 3I/ATLAS.

\section{Results} \label{sec:results}
\subsection{Post-perihelion phase}

\begin{figure*}[ht!]
    \centering
    \includegraphics[width=0.85\linewidth]{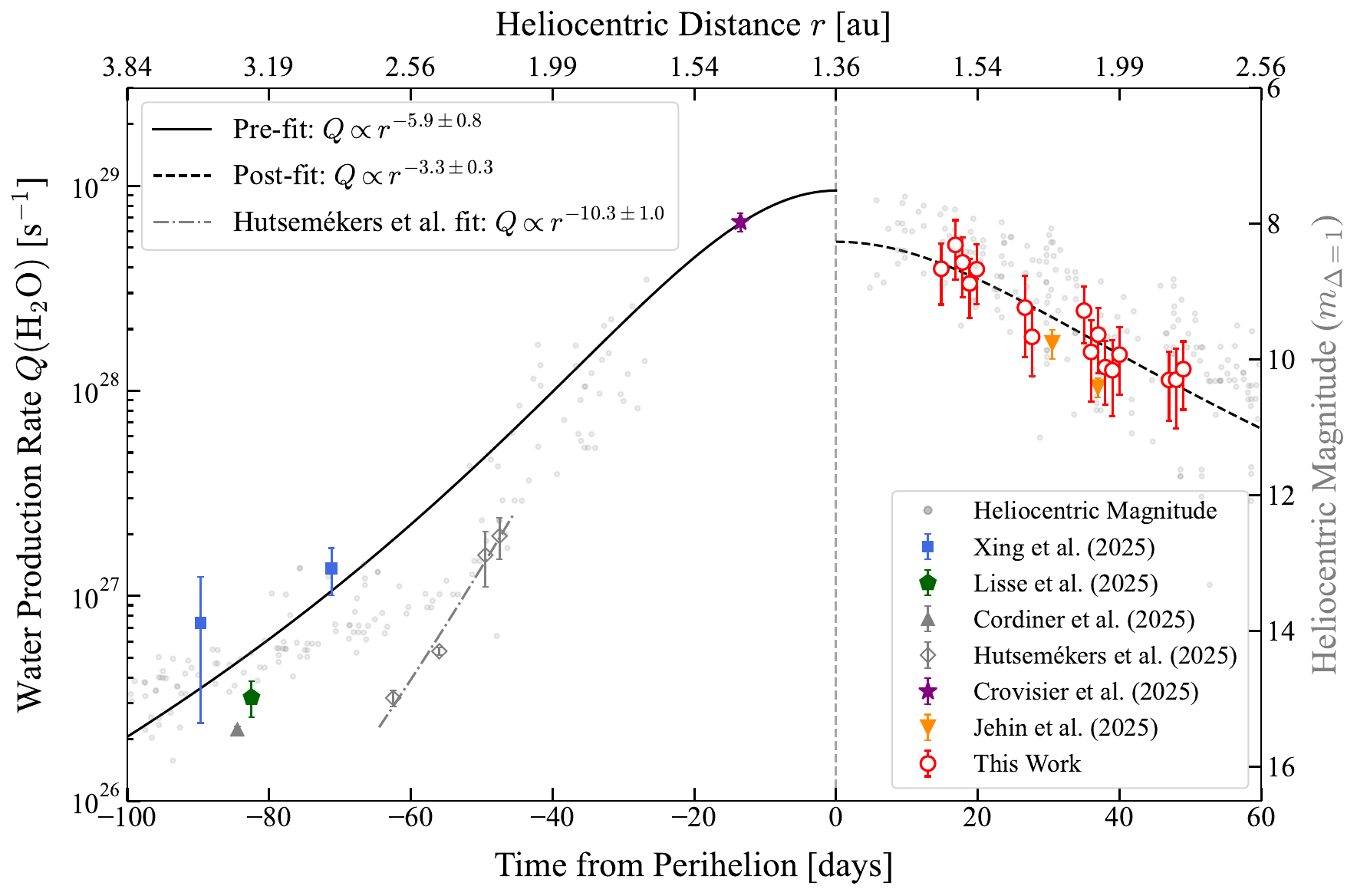}   
    \caption{Heliocentric dependence of the water production rate $Q(\mathrm{H_2O})$ for 3I/ATLAS (left axis, colored symbols), superimposed with the heliocentric magnitude (right axis, grey points). The $Q(\mathrm{H_2O})$ data follow a steep power-law trend pre-perihelion ($n_{\mathrm{pre}} = 5.9 \pm 0.8$), contrasting with a significantly shallower slope post-perihelion ($n_{\mathrm{post}} = 3.3 \pm 0.3$).}
    
    \label{fig:water_prod}    
\end{figure*}

We derived water production rates from SOHO/SWAN observations using the MCPTM described in Section~\ref{subsec:mcptm}. During the post-perihelion phase, the water production rate reached a peak of $Q_{\mathrm{H_2O}} \approx 4 \times 10^{28}$~molecules~s$^{-1}$ (Table~\ref{tab:swan_rates}) and declined steadily with a heliocentric index of $n_{\mathrm{post}} = 3.3 \pm 0.3$, as shown in Figure~\ref{fig:water_prod}. 

\begin{figure*}[ht!]
    \centering
    \includegraphics[width=0.85\linewidth]{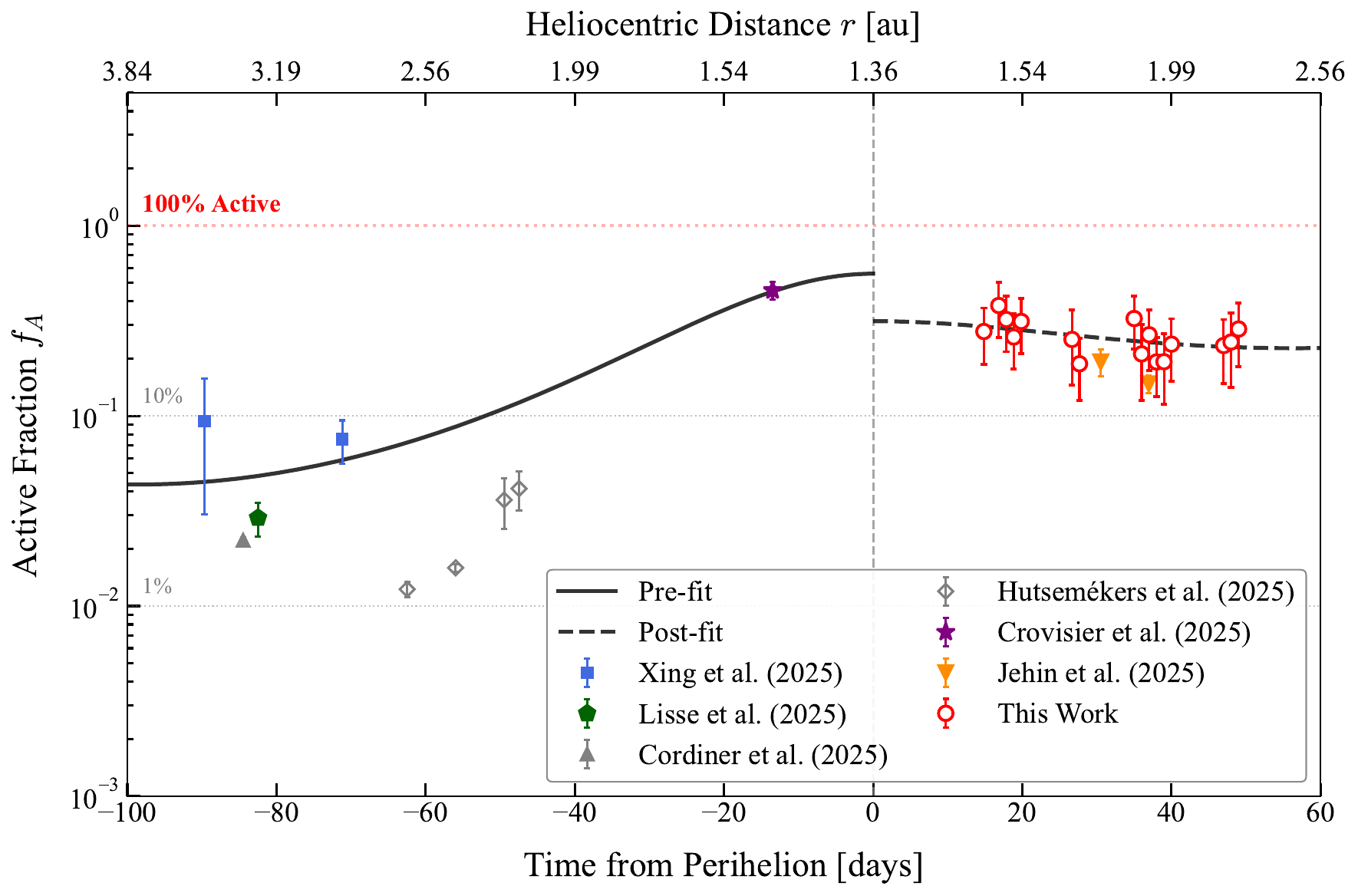}    
    \caption{Variation of the active fraction ($f_{\mathrm{A}}$) with heliocentric distance, calculated assuming a nucleus radius of $R_{\mathrm{N}} = 2.8$~km. The pre- and post-perihelion lines correspond to the power-law fits presented in Figure~\ref{fig:water_prod}. The active fraction reaches a maximum of $\sim 47\%$ pre-perihelion and stabilizes at a constant value of $\sim 30\%$ post-perihelion.}
    
    \label{fig:active_fraction}
\end{figure*}

To quantify the level of activity and interpret the effective active area, we calculated the active fraction $f_{\mathrm{A}}$, following the methodology of \citet{AHearn1995}. This parameter is defined as the ratio of the observed water production rate to the theoretical maximum for a spherical nucleus,

\begin{equation}
f_A(r) = \frac{Q_{\mathrm{H_2O}}(r)}{4\pi R_N^2 \cdot Z_{\mathrm{H_2O}}(r)}
\label{fA}
\end{equation}
where $R_{\mathrm{N}}$ is the nucleus radius. The theoretical specific sublimation rate, $Z_{\mathrm{H_2O}}(r)$, was derived by solving the surface energy balance equation assuming a slow-rotator approximation \citep{Watson1962, Delsemme1971, Cowan1979, Meech2004}:

\begin{equation}
(1-A) \frac{S_{\odot}}{r^2} \cos\theta = \epsilon \sigma T^4 + L(T) Z(T)
\end{equation}
where $S_{\odot}$ is the solar constant. 
We adopted standard thermal parameters for the Bond albedo of $A_{\mathrm{B}} = 0.04$ and the infrared emissivity of $\epsilon = 0.9$ \citep{Meech2004}. By adopting the upper limit of the nucleus radius $R_{\mathrm{N}} = 2.8$~km constrained by HST observations \citep{Jewitt2025a}, we derived the variation of the active fraction with heliocentric distance, as shown in Figure~\ref{fig:active_fraction}.

In the post-perihelion phase, 3I/ATLAS exhibited a nearly constant active fraction of $\sim 30\%$. This corresponds to a mean effective active area of $A_{\mathrm{eff}} \approx 30$~km$^2$, derived via $A_{\mathrm{eff}} = Q_{\mathrm{H_2O}} / Z_{\mathrm{H_2O}}$. This substantial and stable active area indicates that the comet sustained a high level of sublimation-driven activity throughout the observed interval compared with the typical comets in the solar system \citep{Combi2019}.

\subsection{Perihelion Asymmetry} \label{subsec:perihelion_asymmetry}

To investigate the heliocentric dependence of the water production rate, we compiled pre-perihelion datasets from the Neil Gehrels Swift Observatory \citep{Xing2025}, the Nançay Radio Telescope \citep{Crovisier2025}, the SPHEREx mission \citep{Lisse2025}, the James Webb Space Telescope (JWST) \citep{Cordiner2025}, the Very Large Telescope (VLT) \citep{Hutsemekers2025}, and the TRAnsiting Planets and PlanetesImals Small Telescope (TRAPPIST) \citep{Jehin2025a, Jehin2025b}. The water production rates derived from OH observations by SPHEREx and TRAPPIST were standardized following the formalism of \citet{Cochran1993}, while the Nancay radio data were processed according to \citet{Roth2025}.

For the power-law analysis, we restricted our fit to instruments with large projected aperture sizes: SWAN ($\sim 5 \times 10^6$~km), Nançay (beam width $\sim 10^5$ to $10^6$~km), SPHEREx ($\sim 5 \times 10^4$~km), and Swift ($\sim 10^4$ to $10^6$~km). These observations sample a sufficiently large volume of the coma to capture the contribution from distributed sources if any. Conversely, to minimize systematic offsets caused by aperture effects, we excluded the small-aperture measurements from JWST ($\sim 6500$~km) and the VLT ($\sim 800$~km) from the primary power-law fit.

During pre-perihelion, the water production rate followed a relatively steep power-law trend with an index of $n_{\mathrm{pre}} = 5.9 \pm 0.8$. Concurrently, as the comet approached perihelion ($r$ decreasing from $>3$~au to 1.44~au), the active fraction increased from $<10\%$ to a maximum of $\sim 47\%$, indicating a significant increase in the effective active area.

\setcounter{footnote}{0}
Comparison between our post-perihelion measurements and the published pre-perihelion data reveals a significant perihelion asymmetry in water production. The activity profile is characterized by a steep pre-perihelion rise followed by a shallower post-perihelion decline (Figure~\ref{fig:water_prod}). This asymmetry is also evident in the heliocentric light curve obtained from the COBS database\footnote{\url{https://cobs.si}} \citep{Zakrajsek2018}.

\section{Discussion} \label{sec:discussion}

\subsection{Post-perihelion Water Production} \label{subsec:water_post}

During the post-perihelion phase, the water production rate decreased following a power-law trend ($n_{\mathrm{post}} = 3.3 \pm 0.3$), while the active fraction remained constant. In the heliocentric distance range of $1.4 < r < 2.2$~au, the theoretical specific sublimation rate of water ice, $Z_{\mathrm{H_2O}}$ (Eq.~\ref{fA}), scales with a power-law index of $n_{\mathrm{theo}} \approx 2.7$. This is broadly consistent with our observed slope ($n_{\mathrm{post}} \approx n_{\mathrm{theo}}$) given the uncertainties. Consequently, the effective active area ($f_{\mathrm{A}} \propto Q_{\mathrm{H_2O}} / Z_{\mathrm{theo}} \propto r^{-n_{\mathrm{post}}} / r^{-n_{\mathrm{theo}}}$) exhibited relative stability with heliocentric distance. This agreement suggests that the water production was driven predominantly by the varying solar insolation acting on a stable effective active area. 

Our derived water production rates are compatible with those obtained from TRAPPIST observations \citep{Jehin2025a, Jehin2025b}. To ensure internal consistency, we restricted the power-law fitting in this study exclusively to the SWAN dataset.

\subsection{Water Production from the Coma} \label{subsec:coma_water}

The pronounced perihelion asymmetry in the water production suggests a transition in the dominant mechanism of the cometary activity. In contrast to the constant effective active area observed post-perihelion, the steep pre-perihelion rise in water production implies a mechanism that effectively expanded the active surface.

To constrain the underlying mechanisms, characterizing the spatial distribution of the water source is essential. We hypothesize that the sustained water production was dominated by sublimation from a distributed source of icy grains in the extended coma. Direct evidence for the presence of water ice in the coma is provided by the spectroscopic detection of absorption features at $r \approx 4$~au \citep{Yang2025}.

Additional evidence for a distributed source is provided by the aperture dependence of the water production rates. The production rates derived from the small-aperture VLT observations are roughly an order of magnitude lower than the values predicted by the large-aperture pre-perihelion trend at the same heliocentric distance. By comparing the VLT measurements with this extrapolated baseline, we find that the nucleus and near-nucleus coma contribute only a minor fraction ($\sim 10\%$) to the total water production. Consequently, the bulk of the production ($\sim 90\%$) is likely attributable to sublimation from icy grains in the extended coma. A similar aperture effect was observed pre-perihelion at $r \approx 3.3$~au, where the large-aperture Swift measurements yielded higher production rates than the concurrent small-aperture JWST observations. This persistent aperture dependence demonstrates that a substantial fraction of the water production arises from an extended source rather than solely from the nucleus.

In this scenario, icy grains were entrained by super-volatile outflow (e.g., $\mathrm{CO_2}$). These grains remained thermally stable at large heliocentric distances, maintained by the low temperature of the surrounding gas \citep{Sunshine2021} and the intense sublimative cooling derived from the hypervolatiles \citep{Lisse2021}.


As 3I/ATLAS crossed the water-ice sublimation line ($r \approx 2$ to $3$~au), the increasing solar insolation triggered the rapid sublimation of icy grains previously entrained by super-volatiles. Concurrently, water vapor drag progressively became the dominant mechanism for entraining new icy grains into the coma. Pre-perihelion, this transition resulted in a continuous increase in the effective active area (manifested as a rising active fraction), driving both the steep inbound slope in water production and the rapid increase in the optical scattering cross-section. In contrast, the post-perihelion effective active area remained constant. Consequently, the production decline followed the trend of equilibrium sublimation driven solely by the decreasing solar insolation, without the dynamical expansion observed inbound or the depletion of the icy grain population.

The small-aperture VLT measurements reveal a steep increase ($n = 10.3 \pm 1.0$) in the water production rate in the range of $r \approx 2.2$ to $2.6$~au, coincident with a slope discontinuity in the COBS light curve near $r \approx 2.3$~au. 
Restricted to the region sampled by the small aperture, this feature likely marks the onset of rapid water-ice sublimation from the nucleus or the innermost coma.

\subsection{Comparison with 2I/Borisov and Solar System Comets} \label{subsec:comparison_comets}

The pronounced perihelion asymmetry of 3I/ATLAS is analogous to the behavior of the hyperactive Jupiter-family comet 103P/Hartley 2 during its 1997 apparition. \citet{Combi2011a} reported a similarly steep pre-perihelion slope ($n = 6.6 \pm 0.1$) transitioning to a significantly shallower post-perihelion dependence ($n = 3.2 \pm 0.1$). In the case of 103P/Hartley 2, in situ observations by the EPOXI mission demonstrated that its hyperactivity arises from a distributed source of water ice. Specifically, a substantial fraction of the water production is attributed to the sublimation of icy grains entrained into the coma by $\mathrm{CO_2}$ driven jets \citep{AHearn2011}.

We propose a similar mechanism for 3I/ATLAS. Adopting the upper limit of $R_{\mathrm{N}} = 2.8$~km yields a sustained post-perihelion active fraction of $\sim 30\%$ and a pre-perihelion peak of $\sim 47\%$. However, for a smaller nucleus (e.g., $R_{\mathrm{N}} \lesssim 1.9$~km), the derived active fraction would exceed unity ($f_A > 1$). Such values imply that the observed water production cannot be sustained by the nucleus surface alone, requiring a significant contribution from an extended source. This points to a potential hyperactive nature analogous to 103P/Hartley 2, where the activity is dominated by a distributed population of icy grains. 
This scenario is consistent with the evidence for a distributed source inferred from the aperture effects discussed in Section \ref{subsec:coma_water}.

However, the post-perihelion behavior of 3I/ATLAS stands in sharp distinction to that of interstellar comet 2I/Borisov and many dynamically new comets.

The post-perihelion phase of 2I/Borisov was marked by clear evidence of nucleus instability. This period featured optical outbursts attributed to the ejection of large particles and the fragmentation of boulder-sized components \citep{Jewitt2020}. This activity was accompanied by a rapid collapse in the water production rate, which decreased by $\sim$80\% within 20 days, revealing significant surface heterogeneity \citep{Bodewits2020, Xing2020, Yang2021}.

Comparison with dynamically new comets further highlights the stability of 3I/ATLAS. Although C/2009 P1 (Garradd) also exhibited a constant effective active area post-perihelion, its water production peaked significantly early, approximately 100 days prior to perihelion. This pronounced asymmetry is attributed to the chemical heterogeneity of the nucleus, manifested by the premature depletion of an extended halo of icy grains \citep{Combi2013, Feaga2013, Bodewits2014}. In a more extreme example of instability, C/1999 S4 (LINEAR) underwent total disintegration near perihelion, resulting in a precipitous 96\% collapse in water production over an 18-day interval \citep{Makinen2001a}.

In contrast, 3I/ATLAS exhibits a smooth heliocentric evolution in both its water production rate and optical light curve \citep{Eubanks2025, Zhang2025}. This regularity implies a steady sublimation process; the extended source of icy grains was not prematurely depleted. Furthermore, we find no evidence of outbursts, fragmentation events, or significant inhomogeneities in the surface water distribution throughout the perihelion passage.


We note that our analysis, especially regarding the pre-perihelion phase, is limited by the lack of continuous data and the challenges in combining results from different instruments. However, the scenario of a hyperactive nucleus releasing icy grains effectively explains the various photometric, spectroscopic, and water production observations. Due to the potential systematic offsets associated with cross-instrument calibration differences, we cannot entirely rule out other interpretations. Future observations will be crucial to fully understand the activity and evolution of 3I/ATLAS.

\begin{deluxetable}{lcccc}
\tabletypesize{\footnotesize} 
\tablecaption{SOHO/SWAN Water Production Rates of 3I/ATLAS \label{tab:swan_rates}}
\tablewidth{\columnwidth} 
\tablehead{
\colhead{Epoch (Mean Time)} & 
\colhead{$T - T_p$\tablenotemark{a}} & 
\colhead{$r$\tablenotemark{b}} & 
\colhead{$\Delta$\tablenotemark{c}} & 
\colhead{$Q(\mathrm{H_2O})$\tablenotemark{d}} \\
\colhead{(UTC)} & \colhead{(days)} & \colhead{(au)} & \colhead{(au)} & \colhead{($10^{28}$ s$^{-1}$)}
}

\startdata
2025 Nov 13.3 & 14.8 & 1.460 & 2.125 & 3.94 $\pm$ 1.30 \\
2025 Nov 15.3 & 16.8 & 1.488 & 2.099 & 5.15 $\pm$ 1.65 \\
2025 Nov 16.3 & 17.8 & 1.504 & 2.087 & 4.23 $\pm$ 1.37 \\
2025 Nov 17.3 & 18.8 & 1.520 & 2.074 & 3.34 $\pm$ 1.07 \\
2025 Nov 18.3 & 19.8 & 1.537 & 2.061 & 3.92 $\pm$ 1.27 \\
2025 Nov 25.1 & 26.7 & 1.668 & 1.976 & 2.55 $\pm$ 1.08 \\
2025 Nov 26.1 & 27.7 & 1.689 & 1.964 & 1.83 $\pm$ 0.66 \\
2025 Dec 3.5 & 35.0 & 1.862 & 1.884 & 2.46 $\pm$ 0.76 \\
2025 Dec 4.5 & 36.0 & 1.886 & 1.875 & 1.55 $\pm$ 0.66 \\
2025 Dec 5.5 & 37.0 & 1.912 & 1.866 & 1.88 $\pm$ 0.66 \\
2025 Dec 6.5 & 38.0 & 1.937 & 1.857 & 1.30 $\pm$ 0.45 \\
2025 Dec 7.5 & 39.0 & 1.963 & 1.849 & 1.26 $\pm$ 0.51 \\
2025 Dec 8.5 & 40.0 & 1.989 & 1.842 & 1.50 $\pm$ 0.54 \\
2025 Dec 15.5 & 47.0 & 2.180 & 1.807 & 1.13 $\pm$ 0.42 \\
2025 Dec 16.5 & 48.0 & 2.209 & 1.805 & 1.13 $\pm$ 0.48 \\
2025 Dec 17.5 & 49.0 & 2.237 & 1.803 & 1.28 $\pm$ 0.47 \\
\enddata

\tablenotetext{a}{Time relative to perihelion ($T_p =$ 2025 Oct 29.482 UT).}
\tablenotetext{b}{Heliocentric distance.}
\tablenotetext{c}{SOHO-Comet distance.}
\tablenotetext{d}{Water production rate including total uncertainty.}
\end{deluxetable}

\section{Conclusions} \label{sec:conclusions}

In this study, we investigated the heliocentric dependence of the water production rate for interstellar object 3I/ATLAS by combining post-perihelion SOHO/SWAN observations with pre-perihelion multi-instrument measurements. Our primary findings are summarized as follows:

\begin{enumerate}
    
        
    \item The water production rate of 3I/ATLAS exhibited a pronounced perihelion asymmetry, characterized by a steep pre-perihelion rise ($n_{\mathrm{pre}} = 5.9 \pm 0.8$) followed by a shallower post-perihelion decline ($n_{\mathrm{post}} = 3.3 \pm 0.3$). The post-perihelion slope is consistent with the theoretical expectation for equilibrium sublimation of water ice, indicating that the activity was driven primarily by solar insolation acting on a constant effective active area.

    \item 3I/ATLAS exhibits characteristics of a highly active interstellar comet, analogous to the Jupiter-family comet 103P/Hartley 2. We postulate that its water production was dominated by a distributed source of icy grains in the coma.
    
    \item The water production rates of 3I/ATLAS exhibited a smooth, monotonic power-law dependence on heliocentric distance, with no indication of a steep post-perihelion decay. This secular evolution is consistent with the absence of outbursts, macroscopic fragmentation, or rapid fading in concurrent optical monitoring.
    
    
\end{enumerate}



\begin{acknowledgments}

The authors thank the anonymous reviewer and the editor for their helpful and supportive comments.

We gratefully acknowledge the use of data and services provided by the International Astronomical Union's Minor Planet Center (MPC) and the Jet Propulsion Laboratory's (JPL) Horizons system. We acknowledge with thanks the comet observations from the COBS contributed by observers worldwide and used in this research. The results presented in this document rely on data described in \citep{Kretzschmar2018}. These data were accessed via the LASP Interactive Solar Irradiance Datacenter (LISIRD) (\url{https://lasp.colorado.edu/lisird/}).

SOHO is a project of international cooperation between ESA and NASA. We extend a special thanks to the SOHO/SWAN team for maintaining and openly sharing their data with the community over the past 30 years (\url{http://swan.projet.latmos.ipsl.fr/data/L2/FSKMAPS/}). 

For H. Tan, this work carries deep personal significance. From the discovery of his first sungrazing comet in 2009 to the identification of the milestone SOHO-5000, the SOHO mission has played a pivotal role in his life. As the mission approaches its conclusion, we bid farewell to this historic observatory. We remain deeply grateful for its three decades of vigil, paying tribute to the end of an era while looking forward to the future discoveries that will be built upon its foundation.

J.-Y. Li acknowledges the support by the 2024 Xinjiang Autonomous Region Tianchi Talent Program and by Natural Science Foundation of Xinjiang Uygur Autonomous Region No. 2025D01E62.

\end{acknowledgments}


\begin{contribution}
H.T.: Methodology, Software, Formal analysis, Writing – original draft. \\
X.Y.: Methodology, Validation, Visualization, Project administration, Writing – review \& editing. \\
J.-Y.L.: Conceptualization, Supervision, Resources, Writing – review \& editing.
\end{contribution}

\facilities{SOHO}

\software{Astropy \citep{Astropy2013, Astropy2018, Astropy2022}, 
          SciPy \citep{Scipy2020}, 
          Matplotlib \citep{Matplotlib2007}
          }
          

\bibliography{3I_Reference}{}
\bibliographystyle{aasjournalv7}

\appendix
\section{Validation of the Water Production Rate Retrieval Methodology}
\label{sec:appendix_validation}

\begin{figure*}[ht!]
\plotone{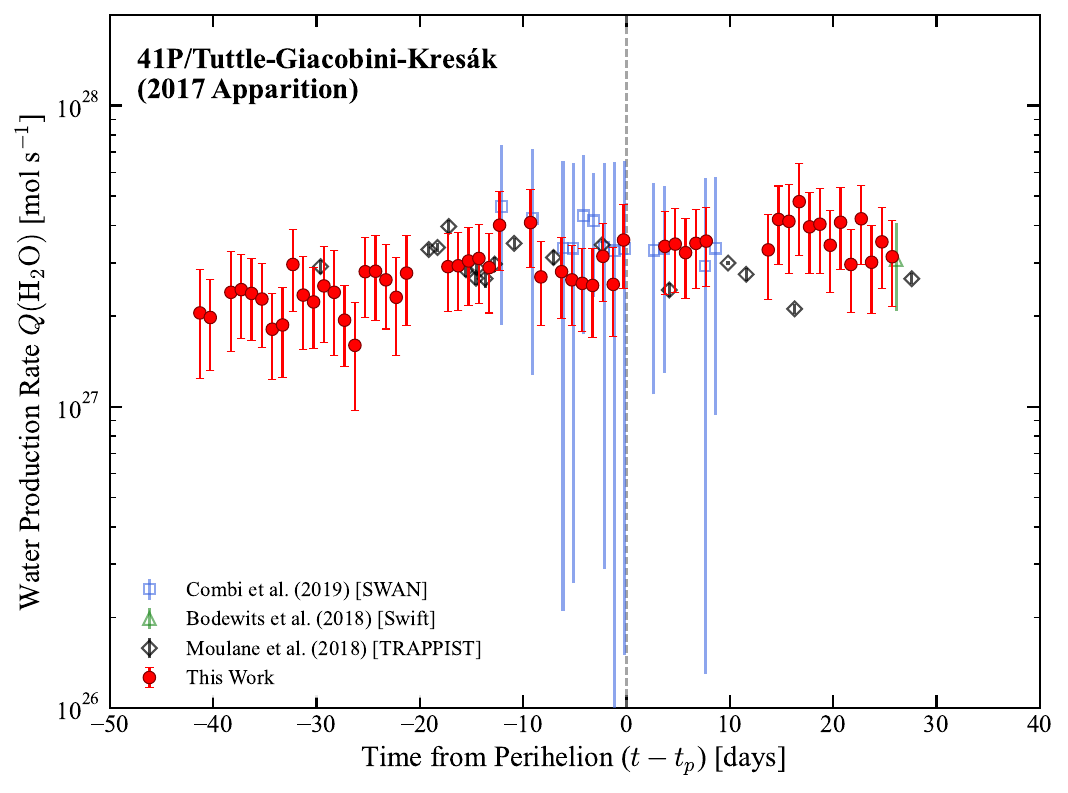}
\caption{Comparison of water production rates for Jupiter-family comet 41P/Tuttle-Giacobini-Kresák derived in this work (red circles) with literature values. The data are plotted alongside the previous SWAN analysis by \citet{Combi2019} (blue squares), TRAPPIST OH observations from \citet{Moulane2018} (black diamonds), and Swift/UVOT measurements \citep{Bodewits2018} (green triangles).
\label{fig:41P_App_Verification}}
\end{figure*}

\begin{figure*}[ht!]
\plotone{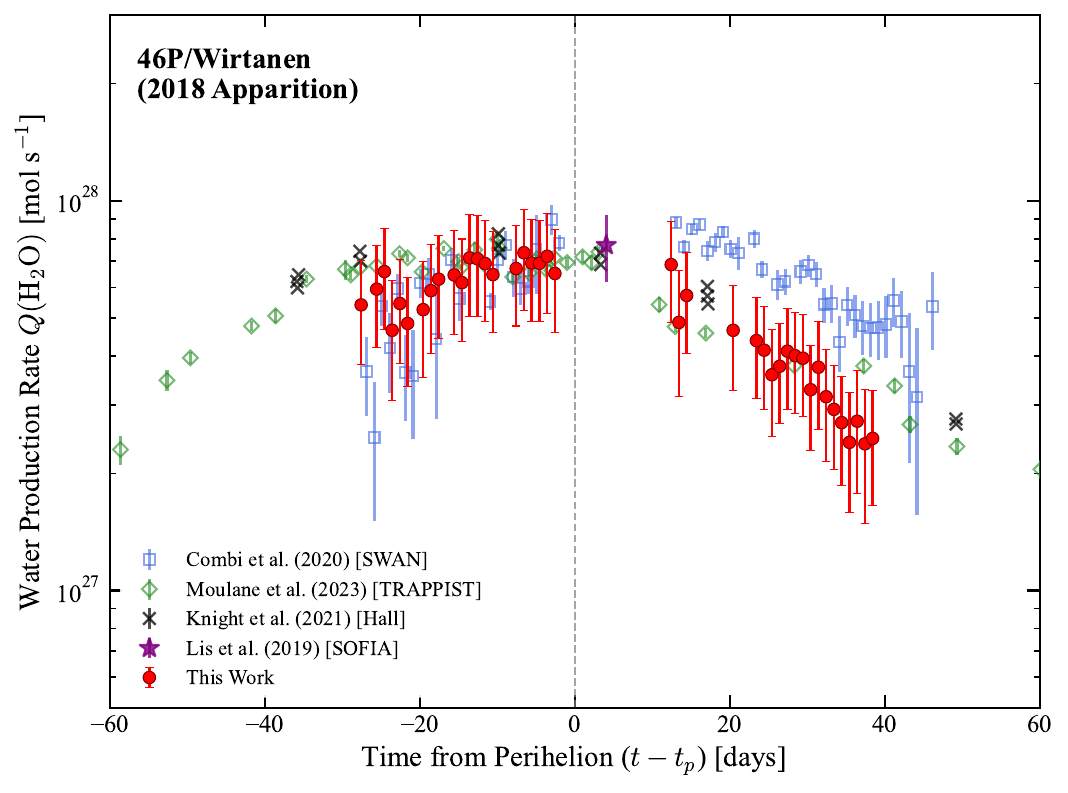}
\caption{Heliocentric evolution of water production rates for 46P/Wirtanen during its 2018 apparition. Rates derived in this work (red circles) are compared with previous SWAN results from \citet{Combi2020} (blue squares), ground-based TRAPPIST photometry \citep{Moulane2023} (green diamonds), and spectroscopic measurements from the Hall telescope \citep{Knight2021} (black asterisks) and SOFIA \citep{Lis2019} (purple star).
\label{fig:46P_App_Verification}}
\end{figure*}

\begin{figure*}[ht!]
\plotone{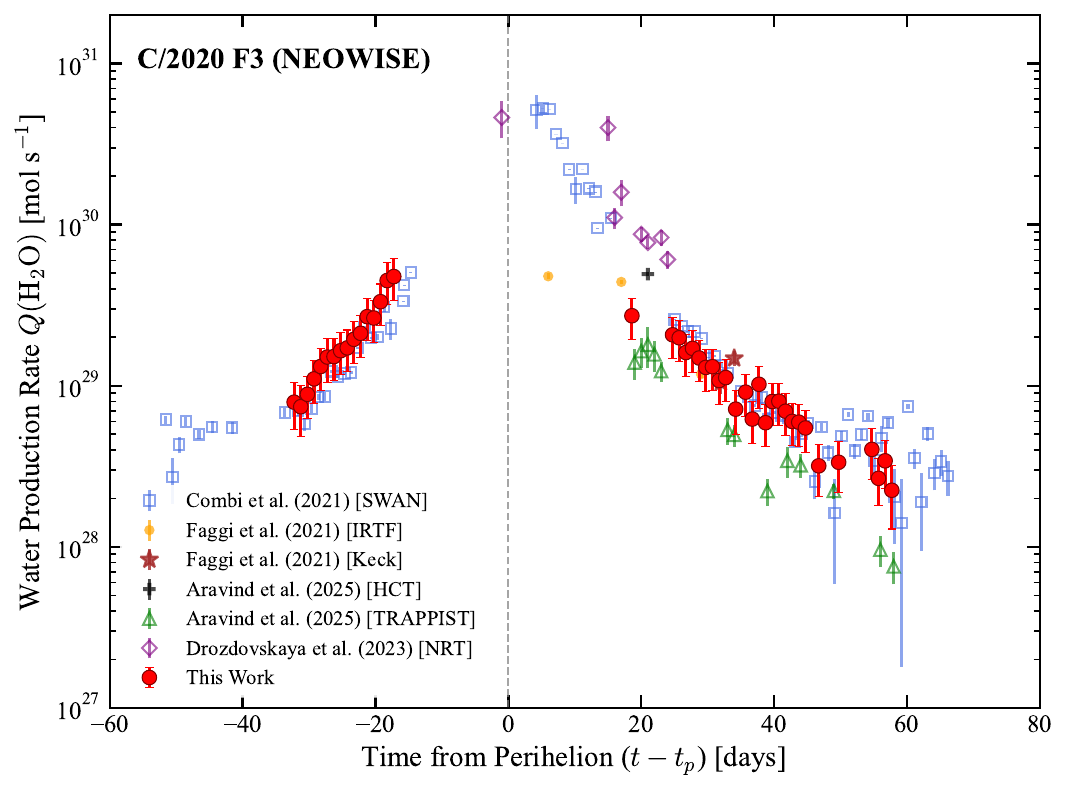}
\caption{Water production rates for the Oort Cloud comet C/2020~F3 (NEOWISE). Results from this work (red circles) are compared with the previous SWAN analysis by \citet{Combi2021} (blue squares), infrared spectroscopy from IRTF (orange circles) and Keck (brown star) \citep{Faggi2021}, OH radio observations from NRT \citep{Drozdovskaya2023} (purple diamonds), and ground-based narrow-band photometry from HCT (black plus) and TRAPPIST (green triangles) \citep{Aravind2025}.
\label{fig:NEOWISE_App_Verification}}
\end{figure*}

To validate our data reduction and modeling pipeline, we benchmarked our results against three well-characterized comets: 41P/Tuttle-Giacobini-Kresák (2017 apparition), 46P/Wirtanen (2018 apparition), and C/2020~F3 (NEOWISE). These targets span nearly three orders of magnitude in water production rate ($10^{27}$ to $10^{30}$~molecules~s$^{-1}$), allowing for a rigorous assessment of our analysis across a wide dynamic range. As illustrated in Figures~\ref{fig:41P_App_Verification}, \ref{fig:46P_App_Verification}, and \ref{fig:NEOWISE_App_Verification}, our derived water production rates show excellent agreement with independent space- and ground-based measurements, confirming the robustness of our methodology.

\textbf{41P/Tuttle-Giacobini-Kresák:} 
As a typical Jupiter-family comet, 41P serves as a benchmark for the sensitivity of our analysis in the low-activity regime ($Q \sim 10^{27}$~molecules~s$^{-1}$). As shown in Figure~\ref{fig:41P_App_Verification}, our derived rates show excellent agreement with both the space-based Swift/UVOT results \citep{Bodewits2018} and ground-based TRAPPIST photometry \citep{Moulane2018}. This concordance between the SWAN Lyman-$\alpha$ (atomic hydrogen) and TRAPPIST OH data validates the photochemical scale lengths and calibration parameters adopted for low-production targets.
\textbf{46P/Wirtanen:} 

Our analysis of 46P validates the robustness of our pipeline in handling the rapidly changing observing geometries during the comet's historic close approach in 2018 December ($\Delta \approx 0.08$~au). As illustrated in Figure~\ref{fig:46P_App_Verification}, our derived water production rates show good agreement with multi-wavelength measurements obtained via high-resolution spectroscopy \citep{Knight2021, Lis2019} and narrow-band photometry \citep{Moulane2023}. Compared to the standard SWAN analysis by \citet{Combi2020}, our re-reduction yields a water production rate with improved temporal consistency.

\textbf{C/2020 F3 (NEOWISE):} 
For the highly active Oort Cloud comet C/2020~F3 (NEOWISE), as shown in Figure~\ref{fig:NEOWISE_App_Verification}, our derived water production rates closely track the heliocentric evolution reported in the standard SWAN analysis by \citet{Combi2021}. Post-perihelion, our results are consistent with independent measurements obtained via infrared spectroscopy (IRTF/Keck; \citealp{Faggi2021}), optical spectroscopy (Himalayan Chandra Telescope (HCT); \citealp{Aravind2025}), and radio observations (Nançay Radio Telescope (NRT); \citealp{Drozdovskaya2023}). Similar to 3I/ATLAS, C/2020~F3 exhibited large heliocentric radial velocities ($|\dot{r}_h| \approx 38$~km~s$^{-1}$). Consequently, accurate correction for the Swings effect, accounting for the Doppler shift of the solar Lyman-$\alpha$ line, was critical for determining precise $g$-factors. The discrepancy with the lower ground-based values \citep{Aravind2025} is consistent with the aperture effects discussed in Section~\ref{subsec:perihelion_asymmetry}, alongside differences in model scale lengths and computation methods identified by those authors.



\end{document}